\documentclass[a4paper,10pt]{article}
\usepackage[utf8]{inputenc}
\usepackage{hyperref}
\usepackage[T1]{fontenc}
\usepackage{amsmath}
\usepackage[symbol]{footmisc}
\usepackage{amsfonts}
\usepackage{setspace}
\usepackage{cite}
\usepackage[dvipsnames]{xcolor}
\spacing{1}
\usepackage{amssymb}
\usepackage{graphicx}
\usepackage[explicit]{titlesec}
\titlespacing\section{0pt}{10pt plus 4pt minus 2pt}{2pt plus 1.5pt minus 1.5pt}
\usepackage{float}
\usepackage{lipsum}
\usepackage{multicol, blindtext, caption}
\usepackage{xcolor}
\usepackage[font=footnotesize,labelfont=bf, skip=10pt]{caption}
\definecolor{azul}{rgb}{0, 0, 0}
\usepackage[left=1.5cm, right=1.5cm, top=2.0cm, bottom=2.5cm]{geometry}
\usepackage{hyperref}

\usepackage[normalem]{ulem}

\newcommand{\unit}[1]{~\mathrm{#1}}

\begin{document}

	\begin{center}
		{\Large \textbf{Influence of carrier density and disorder on the Quantum Hall plateau widths in epitaxial graphene}}\\
		\vspace{5mm}
		{\large Ignacio Figueruelo-Campanero$^{1,2}$\footnote[1]{ignacio.figueruelo@imdea.org}, Yuriko Baba$^{3,4}$\footnote[1]{yuriko.baba@uam.es}, Alejandro Jimeno-Pozo$^1$, Julia García-Pérez$^{1}$, Elvira M. González$^{1,2}$, Rodolfo Miranda$^{1,5,6}$, Francisco Guinea$^{1,7}$, Enrique Cánovas$^{1}$, Daniel Granados$^{1}$, Pierre Pantaleón$^{1}$, Pablo Burset$^{3,4,6}$ and Mariela Menghini$^{1}$\footnote[1]{mariela.menghini@imdea.org}  }\\
		\vspace{3mm}
		$^1$\textit{IMDEA Nanociencia, Cantoblanco, 28049 Madrid, Spain}\\
		$^2$\textit{Facultad Ciencias Físicas, Universidad Complutense de Madrid, 28040 Madrid, Spain}\\
        $^3$\textit{Departamento~de~Física~Teórica~de~la~Materia~Condensada\char`,~Universidad Autónoma~de~Madrid\char`,~28049~Madrid\char`,~Spain}\\
        $^4$\textit{Condensed Matter Physics center (IFIMAC), Universidad Autónoma de Madrid, 28049 Madrid, Spain}\\
        $^5$\textit{Departamento~de~Física~de~la~Materia~Condensada\char`,~Universidad Autónoma~de~Madrid\char`,~28049~Madrid\char`,~Spain}\\
        $^6$\textit{Instituto Nicolás Cabrera, Universidad Autónoma de Madrid, 28049 Madrid, Spain}\\
        $^7$\textit{Donostia International Physics Center, Paseo Manuel de Lardizábal 4, 20018 San Sebastián, Spain}\\      
	\end{center}

Since its discovery, graphene has been one of the most prominent 2D materials due to its unique properties and broad range of possible applications. In particular, the half-integer Quantum Hall Effect (HI-QHE) characterized by the quantization of Hall resistivity as a function of  applied magnetic field, offers opportunities for advancements in quantum metrology and the understanding of topological quantum states in this 2D material. While the role of disorder in stabilizing quantum Hall plateaus (QHPs) is widely recognized, the precise interplay between the plateaus width, disorder, mobility and carrier density remains less explored.
In this work, we investigate the width of the $\nu=6$ QHP in epitaxial graphene Hall bars, focusing on two distinct regions of the device with markedly different electronic mobilities. Depending on the storage conditions, it is possible to modify the carrier density of graphene QHE devices and consequently increase or reduce the mobility. Our experiments reveal mobility variations of up to 200$\%$ from their initial value. In particular, the sample storage time and ambient conditions cause also noticeable changes in the positions and extension of the QHPs. Our results show that the QHP extension for $\nu=6$ differs significantly between the two regions, influenced by both mobility and disorder, rather than solely by carrier density. Transport simulations based on the Landauer-Büttiker formalism with Anderson disorder in a scaled model reveal the critical role of impurities in shaping graphene transport properties defining the extension of the QHPs. This study provides valuable insights into the interplay between mobility, disorder, and quantum transport in graphene systems.
\\\\
		\vspace{2mm}
		\underline{\textbf{Keywords:}} \hspace{2mm} \textbf{ Epitaxial Graphene, Quantum Hall Effect, Disorder, Quantum transport simulations}

\begin{multicols}{2}
\section*{I. Introduction} \label{Intro}

\begin{figure*}
    \centering
    \includegraphics[width=0.9\textwidth]{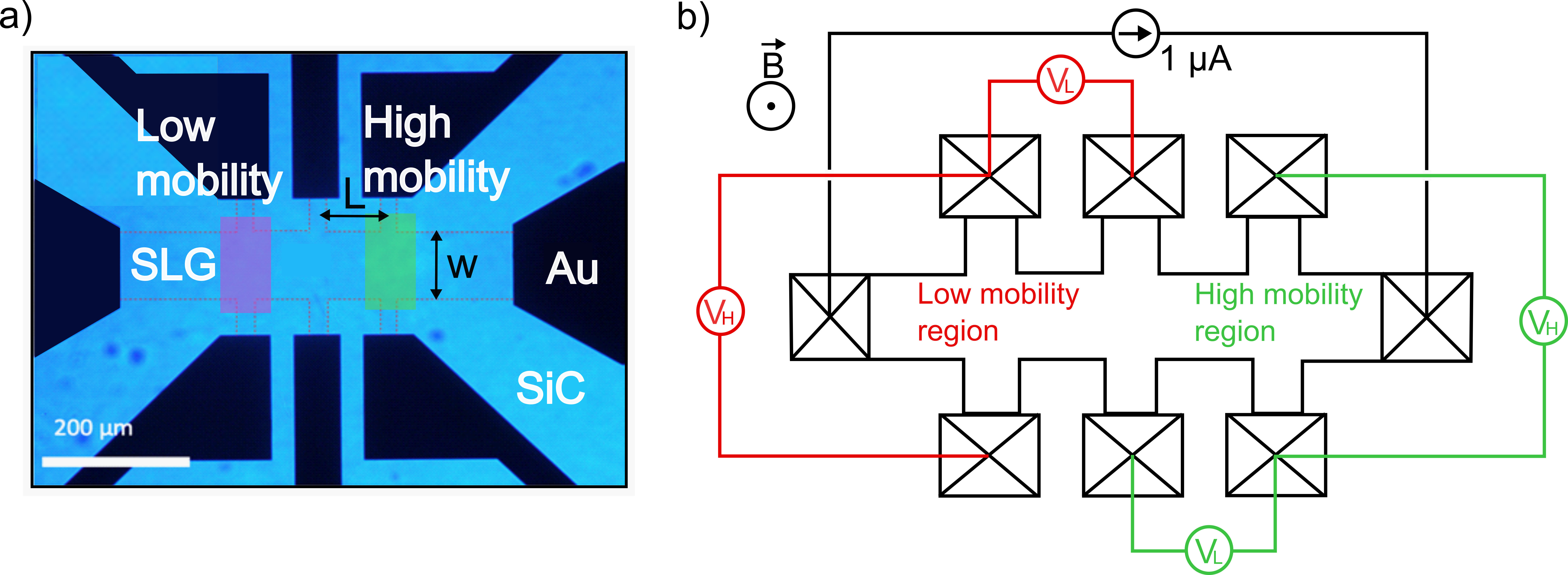}    \caption{a) Optical microscope image of a $W = L =100$ $\mu\mathrm{m}$ graphene Hall bridge. The SLG is indicated by the dotted area, blue contrast corresponds to SiC substrate and black regions correspond to Ti/Au (10/90 nm) contacts. The high mobility region corresponds to the green shaded area in the SLG, and low mobility corresponds to red one. b) Schematics of the configuration used for the longitudinal ($V_L$) and Hall ($V_H$) voltage measurements presented in this work. Green (red) color corresponds to the configuration used to measure the high (low) mobility region.}
    \label{Fig1}
\end{figure*}
Over the last decade, graphene has been one of the most widely studied 2D material in condensed matter physics due to its unique properties and broad range of applications \cite{Novoselov_2005, Castro_Neto_2009, Geim_2007, Novoselov_2012}. In particular, the relativistic electronic nature and its associated Berry phase, together with the sub-lattice degree of freedom of graphene lead to the appearance of Quantum Hall Plateaus (QHPs) at particular Hall resistivity values according to the formula $\rho_{xy}=h/[4e^2(N+1/2)]= h/(\nu e^2)$, for $N=0,1,\dots$ or $\nu=2,6,\dots$ This quantization leads to the so-called half-integer Quantum Hall Effect (HI-QHE) \cite{Novoselov_2007, Zhang_2005, Novoselov_2005}, since not all of the plateaus observed in non-relativistic 2D electron gases (2DEGs) can be stabilized in the case of graphene. The possibility to induce this quantum state in graphene opens a plethora of possibilities not only as a new platform to study the QHE or related phenomena such as fractional QHE \cite{Lu_2024, Bolotin_2009} or Quantum Spin Hall Effect \cite{Kane_2005}, but also in the field of quantum metrology for the development of a Quantum Resistance Standard (QRS) based on graphene \cite{Cage_1985, Yin_2024}. 

\noindent
The QHE has been extensively studied since its discovery, leading to promising research areas such as the role of \textit{k-space} topology in condensed matter systems \cite{Avron_2003, Hatsugai_1997}. However, many aspects of this macroscopic quantum state remain still open questions in literature as there are no theoretical models that can accurately predict the experimental characteristics of QHE at finite temperatures. In this sense, an important aspect of the QHE is the extension or width of the QHPs in magnetic field. There are a few works \cite{Groshev_1994, Kawaji_1993, Schurr_2004, Yi-Thomas_2025} devoted to the study of the plateau width and most of them are focused on 2DEG in AlGaAs/GaAs heterostructures. It has been argued that the QHP width could be related to different transport parameters such as the mobility, the quantum lifetime or the distribution of impurities \cite{Groshev_1994, Kawaji_1993, Schurr_2004}; however, none of them are able to explain the nature of the QHP widths in terms of parameters such as carrier density, mobilities or disorder in the samples. It is generally accepted that disorder is necessary to stabilize QHPs. In order to observe the plateaus, the Fermi level needs to get pinned in the mobility gap between Landau levels (LLs), where only the edge channels allow ballistic carrier transport \cite{Datta_1995}. The Fermi level is pinned when the density of (localized) states in the mobility gap is non-negligible. In large samples, in the absence of disorder, the Fermi level would always be close to the delocalized LLs where the density of states is maximum, leading to no observable QHPs. Nevertheless, there are no bounds, neither theoretically nor experimentally, on how  disorder affects the QHPs stabilization and width  as it is hard or even impossible to control disorder experimentally. It has also been theorized that QHPs could also be present in the absence of disorder due to the formation of a quasi-particle Wigner crystal whenever the Fermi level is close to integer filling factors \cite{Groshev_1994, Kim_2021}. 

\noindent
In this work, we focus on the electronic transport properties and the QHE characteristics in an epitaxial graphene Hall bar. In particular, we have tracked the changes with time of the $\nu=6$ QHP width in two different device regions with different mobilities for a range of carrier densities. By absorption/desorption of mostly oxygen species on the surface via different device storage conditions, we were able to modify the carrier density of the sample without the need of a gate voltage. Carrier density modulation by absorption/desorption can modify the mobility of graphene up to 200$\%$ from its initial value. These carrier density modifications are also reflected in the QHE state, affecting, in particular, the extension of the $\nu=6$ QHPs. 
 We have analyzed theoretically the effect of impurities and disorder in the transport properties of a graphene Hall bar and their implications in determining the QHPs width. We have performed transport calculations within the Landauer-Büttiker formalism in tight-binding model \cite{Groth_2014} including Anderson disorder. 
Our results show that the observed  QHP width variations cannot be attributed only to the changes in carrier density but are also related to the disorder present in the sample as reflected in the difference in mobility between the two studied device regions. The observed features in the QHPs are qualitatively captured by the theoretical calculations in graphene  with localized disorder.  

\noindent
This study provides a detailed examination of the interplay between mobility, disorder, and carrier density in stabilizing the QHPs in epitaxial graphene. By leveraging oxygen specie absorption/desorption to modulate carrier density in regions with distinct mobilities, we uncover how these factors collectively influence the width of the $\nu=6$ QHP. The use of transport simulations with Anderson disorder offers valuable insights into the microscopic mechanisms governing quantum transport. These findings pave the way for future research into controlling QHPs through tailored disorder and mobility engineering, with potential implications for quantum device optimization and metrology applications.

\section*{II. Experimental methods} \label{Exp}
\begin{figure*}
    \centering
    \includegraphics[width=1\textwidth]{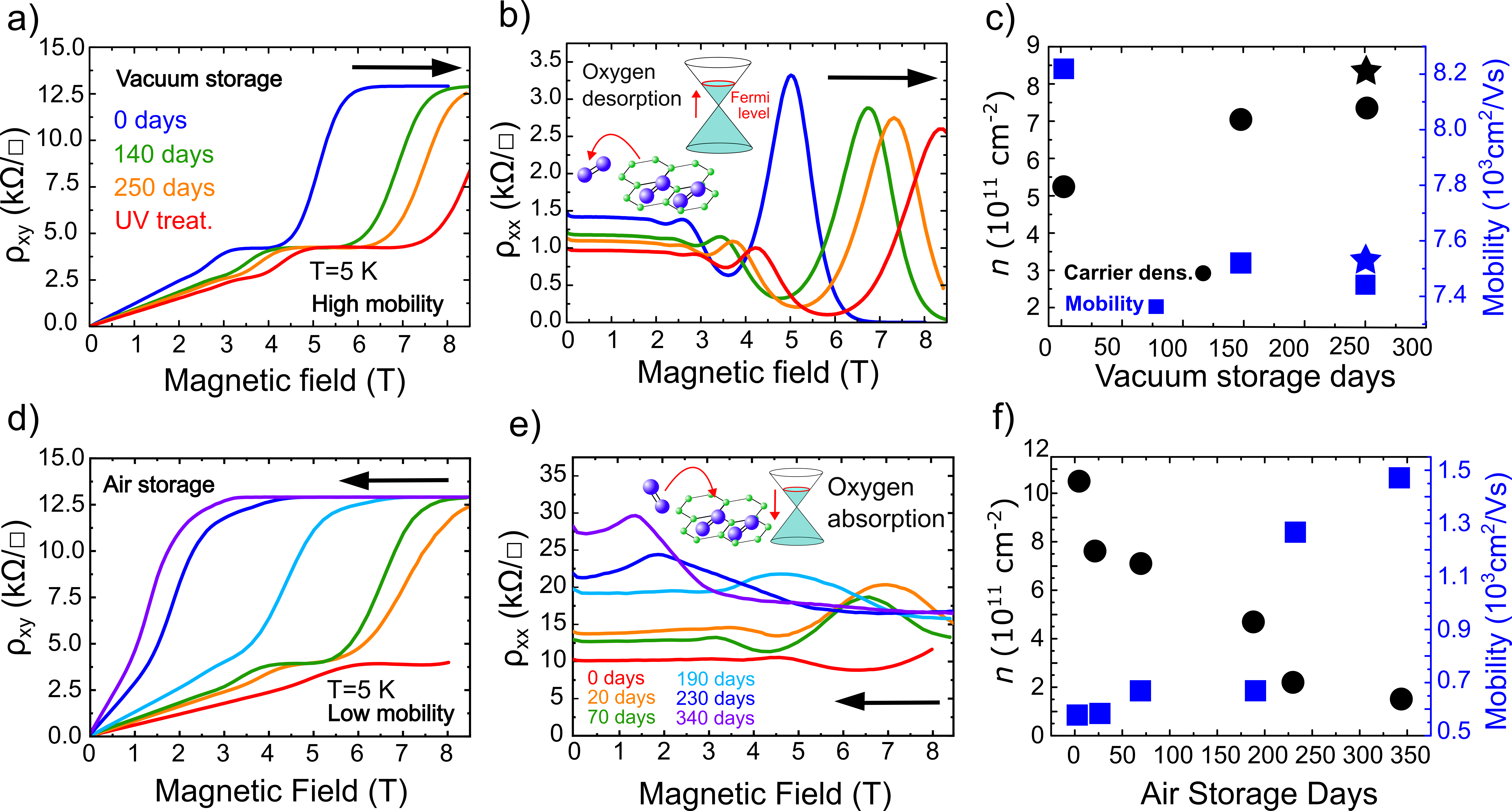}    \caption{a) Hall resistivity and b) longitudinal resistivity as a function of the applied magnetic field for different vacuum storage times and for the UV treatment (red line) at 5 K. c) Evolution of carrier density (black dots) and mobility (blue dots) with vacuum storage time. Star shape markers correspond to UV treatment ($387 \unit{nm}$, $5\unit{s}$, $60 \unit{W/cm}^{-2}$). d) Hall resistivity and e) longitudinal resistivity as a function of the magnetic field for different air storage times at $5 \unit{K}$. f) Evolution of carrier density (black dots) and mobility (blue dots) with air storage time. The black arrows in a), b), d) and e) point the direction of the shift in the Hall plateaus due to aging in each storage condition.}
    \label{Fig2}
\end{figure*}
A commercial single layer graphene (SLG) from Graphensic AB epitaxially grown on 4H-SiC (0001) substrate was used to fabricate Hall bridges in a 2-step standard optical lithography process. First, graphene Hall bars were defined by an optical lithography followed by reactive ion etching in Ar plasma. After that, a second lithographic process was done to fabricate the electrical contacts by a subsequent evaporation of 10/90 nm of Ti/Au using an e-beam evaporator. Side contacts were realized to reduce the capacitance effects and contact resistance at the graphene/Ti/Au interface \cite{Giubileo_2017}. The measurements presented in this manuscript were performed on a device with 100 $\mu$m channel width ($W$) and 100 $\mu$m separation between voltage pads ($L$), see Fig.~\ref{Fig1}a (note that for the graphene Hall bridge $L = W$ implies that the resistance is equal to the resistivity, i.e., $R_{xx} = \rho_{xx}$). The Hall device was bonded to a chip carrier and mounted in a $^4$He flow cryostat that allows temperature control down to $1.5$~K and is equipped with a $9 \unit{T}$ superconducting coil. Magneto-transport measurements using a 4-probe configuration were performed in different regions of the device (see red and green areas in Fig.~\ref{Fig1}a). The longitudinal and Hall (transversal) resistivities were obtained simultaneously using a Keithley 6221 current source and a 2-channel Keithley 2182A nanovoltmeter. A scheme of the measurement configuration is presented in Fig.~\ref{Fig1}b. All results presented in this work were obtained measuring in Delta mode with a current of 1 $\mu$A. 

\noindent
Numerical calculations of the transport properties of a graphene Hall bar were performed considering a tight-binding model with nearest neighbors hopping (see Section 1 of the Supplementary Information for more details on the model employed). By including onsite Anderson disorder with strength $\omega_A$, we studied the effect of impurities on the longitudinal and transversal graphene resistance. The resistance was obtained via the conductance matrix calculated within the Landauer-Büttiker framework employing Kwant \cite{Groth_2014}.

\section*{III. Results $\&$ discussion} \label{Res}
\subsection*{1. Carrier density modulation and the QHE under vacuum and air storage}

The longitudinal and Hall resistances of graphene were characterized as a function of the applied magnetic field at different temperatures. The results at room temperature show one order of magnitude difference in the electron mobilities between the green and red regions marked in Fig.~\ref{Fig1}a, despite both being subjected to the exact same fabrication conditions. To study graphene’s electrical properties for different carrier concentrations, we analyzed the evolution of the longitudinal and Hall resistances under different storage conditions in the two different mobility regions, High$_\text{mob}$ (green) and Low$_\text{mob}$ (red) (Fig.~\ref{Fig1}a). As reported in previous studies \cite{Lara_Avila_2011, Bagsican_2017, Li_2019, Ryu_2010, Lartsev_2014, Yin_2022, Tian_2010, He_2018}, the carrier density in epitaxial graphene can be tuned by adsorption or desorption of different types of molecules at the surface. 

\noindent
Adsorbed oxygen among other molecules present in the atmosphere can lead to a change in electron carrier density as they act as electron acceptors decreasing the electron carrier concentration \cite{Bagsican_2017, Ryu_2010}. The results for High$_\text{mob}$ (Low$_\text{mob}$) region as a function of vacuum storage (air storage) are presented in Fig.~\ref{Fig2} top (bottom) panels. It is important to remark that in our graphene Hall bar the transport is dominated by electrons for all the data shown in the manuscript. 

\noindent
The graphene electronic properties are analyzed by extracting the carrier density and mobility as a function of temperature. First, the carrier density is obtained from the Hall resistivity, $\rho_{xy}$, at low magnetic fields, Fig.~\ref{Fig2}a, following the Drude formalism $n=(d \rho_{xy}/d B)^{-1}/e$, where $e$ is the electron charge and $B$ is the applied magnetic field. For the High$_\text{mob}$ region, the as-fabricated sample presents a carrier density of $5.1\cdot 10^{11}\unit{cm}^{-2}$ at 5 K (see Supplementary Information Section 2 for temperature dependences). 

\noindent
The slope $d\rho_{xy}/dB$ at low magnetic fields decreases with storage time, meaning carrier density increases when the sample is stored in vacuum. The increase in $n$ observed in our experiments can be explained in terms of removal of adsorbed molecules from the surface of graphene as the sample is kept more time in vacuum conditions. To confirm that the carrier modulation is mainly driven by oxygen molecules, UV laser annealing was done to remove possible molecules still present at the surface of graphene after 250 days of storage in vacuum \cite{Wang_2017}. We illuminated the sample with an UV laser ($387 \unit{nm}$) during $5 \unit{s}$ with a fluence of $60 \unit{W/cm}^2$ to minimize heating effects. Immediately after the UV treatment, the sample was mounted in the cryostat and measured. An increment in the carrier density of $\Delta n =10^{11}\unit{cm}^{-2}$ was observed (star shape marker in Fig.~\ref{Fig2}c), suggesting that the modulation of carrier density for vacuum storage comes mostly from desorption of molecular oxygen. 

\noindent
The electron mobility is extracted following the Drude formula $\mu=(L/W)[en\rho_{xx} (B = 0)]^{-1}$, using the carrier density previously obtained from the Hall resistivity slope and the values of the longitudinal resistivity at zero magnetic field $\rho_{xx}$ (0), Fig.~\ref{Fig2}b. The as-fabricated sample has a mobility of $8500 \unit{cm^2/Vs}$ in accordance with other epitaxial graphene reported in literature \cite{Lartsev_2014}. We observe that as carrier density increases with vacuum storage time, this has a detrimental effect on the mobility, which decreases over time reaching  87$\%$ of the starting value after 250 days. This effect can be explained by considering that, as the Fermi level is increased, the available number of states for an electron to scatter is larger, thus having a detrimental effect on the mobility. In contrast, an increment of $\Delta \mu \approx 50 \unit{cm^2/Vs}$ in the mobility after the UV treatment was observed. In this case, the increase in carrier density accompanied by an increase in mobility could be attributed to thermal annealing effects due to laser heating. 

\noindent
To summarize, a carrier density increase and a mobility reduction are observed with vacuum storage time. Apparently, carrier density dictates the evolution of mobility in the sample \cite{Hwang_2007, Nomura_2007, Turyanska_2017, Makarovsky_2017}. In this case, when more oxygen adatoms are desorbed over time, more carriers are present in the sample and the overall effect is lowering the mobility. This means that oxygen adatoms only play the role of modulating the carrier density and have a minor effect in the scattering rates of the system. This is reasonable as molecular oxygen on graphene should be weakly bound via physisorption \cite{Docherty_2012, Schedin_2007} and thus, minimal changes are induced directly on the graphene lattice. 

\noindent
For high enough magnetic fields, the system enters the Quantum Hall regime. In the as-fabricated sample, we observe the appearance of plateaus $\nu=6$  ($N=1$) around $3\unit{T}$ and $\nu=2$ ($N=0$) around $6\unit{T}$, and their corresponding Shubnikov de Haas (SdH) oscillations in the longitudinal resistivity $\rho_{xx}$, see Fig.~\ref{Fig2}b. The effect of removing oxygen from the surface via vacuum storage is also observed in the location and extension of QHPs and SdH oscillations. As storage time increases (and thus carrier density), the position of SdH oscillations is shifted towards higher magnetic fields. This is expected as carrier density is increasing and therefore a higher magnetic field is required to accommodate the gained extra carriers (electrons) due to increased degeneracy of the LLs (see Supplementary Information Section 3). For the case of the UV treatment, the QHP $\nu=2$ appears at magnetic fields beyond the largest magnetic field available in our setup (in agreement with a further increase in the carrier density). However, the plateau $\nu=6$ is still visible and even a glimpse of the $\nu=10$ can be discerned. 

\noindent
In the Low$_\text{mob}$ region (red area in Fig.~\ref{Fig1}a), the initial carrier density was of the order of $10^{12} \unit{cm}^{-2}$ and the mobility $640 \unit{cm^2/Vs}$. Notably, the QHE was also stabilized in this region despite the difference in mobility of one order of magnitude, see Fig.~\ref{Fig2}d. In this case, the high initial electron carrier concentration (red curve) does not allow the observation of the $\nu=2$ plateau in the magnetic field range accessible in our experiments but the plateau $\nu=6$ is located at a similar field as to the last measurement presented for the High$_\text{mob}$ region (250 days in vacuum + UV). For this part of the study, we followed the evolution of graphene transport properties as the sample was kept in air. As time passed, the electron density decreased due to the air exposure and consequently re-adsorption of oxygen molecules. An increase of the longitudinal resistance at zero magnetic field is also observed in these conditions, see Fig.~\ref{Fig2}e. 

\noindent
The evolution of carrier density and mobility for this region as a function of time is summarized in Fig.~\ref{Fig2}f. During air storage, the evolution of carrier density and mobility retraces back the behavior seen in the previous case. The exposure to air allows to fully recover the initial carrier concentration present originally in the High$_\text{mob}$ region and to go even below that value down to $3\cdot 10^{11} \unit{cm}^{-2}$. 
After 190 days in air, the plateau $\nu=2$ is established within the accessible field range, and after $\sim$340 days the quantization is observed by only applying $3 \unit{T}$ (see Fig.~\ref{Fig2}d). The corresponding longitudinal resistivities for the different storage times depicted in Fig.~\ref{Fig2}e show some modulation with the magnetic field corresponding to the SdH oscillations. Nevertheless, none of them shows the distinctive minima near zero resistance as it was observed in the High$_\text{mob}$ region when entering the $\nu=2$ QHP regime. The longitudinal resistivity in the QHE evolves accordingly to the evolution of the density of states at the Fermi level with the applied magnetic field. Clean samples host narrow LLs when a magnetic field is applied. In this regime, backscattering is suppressed between edge channels and consequently, the longitudinal resistivity presents marked drops near zero resistance when the Fermi level is located between the LLs. When impurities are added to the system, the LLs start to widen, increasing the number of localized states in the mobility gap. At a certain point, this can give rise to a finite longitudinal conduction by means of thermally activated or hopping transport mechanisms, thus reducing the expected quantized conductivity \cite{Laughlin_1981}.

\subsection*{2.	Role of mobility, carrier density and scattering rate in the QHPs }
\begin{figure*}
    \centering
    \includegraphics[width=0.9\textwidth]{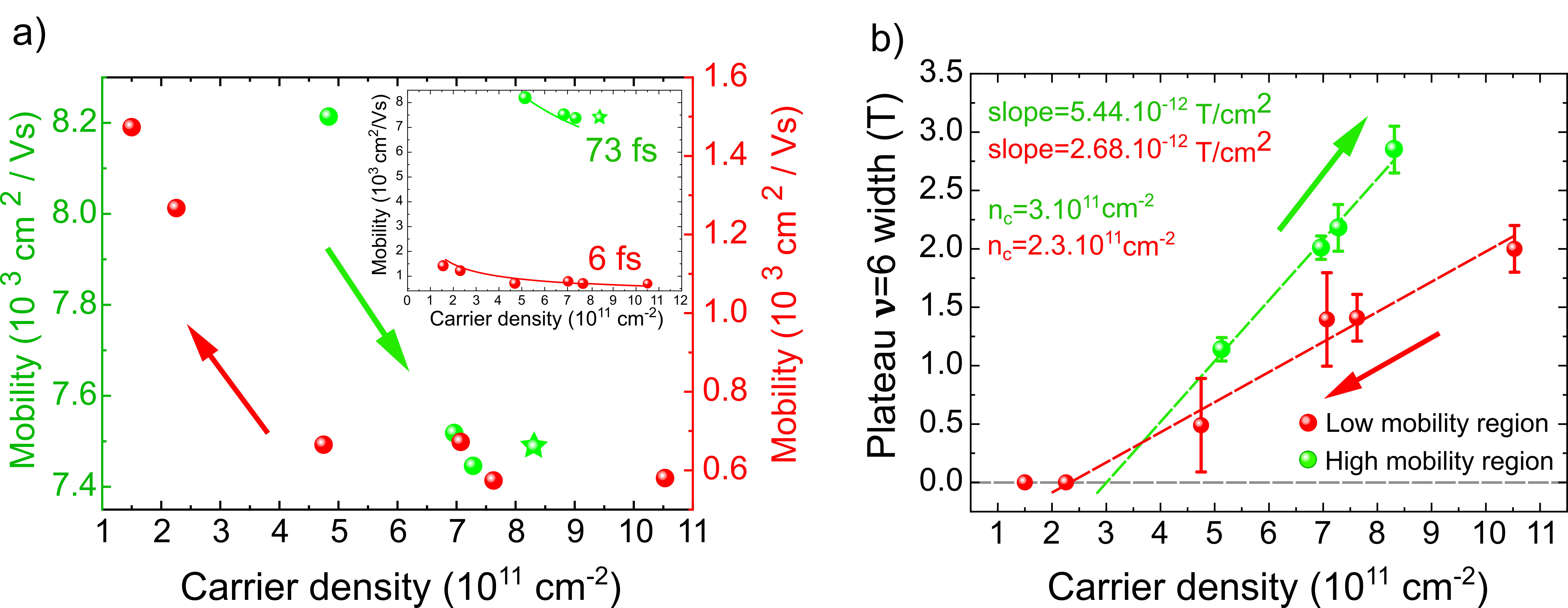}    \caption{a) Characteristic mobility as a function of carrier density for the high mobility region (left axis) and the low mobility region (right axis). Inset: Same as main panel but using the same scale of mobility for both regions. The continuous lines correspond to a fitting using the model described in \cite{Shon_1998}, see text. b)  QHP width for $\nu=6$ for different carrier densities for the high and low mobility regions. In both figures, the arrows indicate the direction of evolution of carrier density with time in each region. Dotted lines represent the linear fit for each region with values for the slope and intercept ($n_c$).}
    \label{Fig3}
\end{figure*}
We now discuss the relation between the carrier density, mobility and scattering rates for the results presented in the previous section and their influence in the QHP extension. 

\noindent
Figure~\ref{Fig3}a shows the behavior of mobility as a function of the carrier density for both high and low mobility regions. As previously mentioned, mobility seems to be modulated mostly by the carrier density. In both regions, mobility decreases as the carrier density increases. This trend has been reported in previous works and is expected according to different theoretical frameworks \cite{Turyanska_2017, Makarovsky_2017, Gosling_2021}. Classical Drude’s model renders for graphene the relation $\mu(n)=\tau ev_F/(\sqrt{n}\pi \hbar)$, where $v_F$ is the Fermi velocity and $\tau$ the average scattering rate, left as a constant parameter. Beyond this model, by using Boltzmann theory with Born approximation, it is possible to obtain the scattering rate dependence with carrier density $\tau(n)$ \cite{Gosling_2021}. This dependence now results in a more general expression for mobility vs carrier density as $\mu(n)\propto n^\beta$, where the exponent $\beta$ depends on the Fermi level position, type of scatterer, and the density ($n_\text{imp}$) and strength ($U_0$) of impurities \cite{Gosling_2021, Shon_1998}. In our case, we found that the exponents $\beta(\text{High}_\text{mob}) =-0.30$ and $\beta(\text{Low}_\text{mob})=-0.57$ fit reasonably well our experimental data, see inset of Fig.~\ref{Fig3}a. 

\noindent
For each region, an average scattering rate is extracted using Drude’s formula for mobility. We obtain a scattering rate of $73\unit{fs}$ for the high mobility region and $6 \unit{fs}$ for the low mobility one, i.e., one order of magnitude difference as obtained for the mobilities (inset Fig.~\ref{Fig3}a). A generalized relation between impurity density $n_\text{imp}$ and strength, $U_0$, can be deduced from the extended Drude model of Ref. \cite{Gosling_2021} for the scattering rate $\tau \propto n_\text{imp}^{-1} (A_{sr} U_0^2)^{-1}$, where $A_{sr}$ is the effective cross-section of defects. This formula implies that the effects of $n_\text{imp}$ and $U_0$ are mixed on the resulting scattering rate and, thus, in the mobility. As a result, it is not experimentally possible to disentangle the nature of the observed mobility differences in the studied sample. However, for our case it is reasonable to consider that the type of scatterers present in both regions of the sample is similar and thus have comparable scattering strengths, $U_0|_\text{Low} \approx U_0|_\text{High}$. Consequently, the observed difference in mobilities can originate mainly from a difference in impurity density between each region. We can extract the following relation for the product between impurities density and strength in both regions: $\tau_\text{High}/ \tau_\text{Low} \propto (n_\text{imp} U_0^2|_\text{Low})/(n_\text{imp} U_0^2|_\text{High}) \approx 12$ (see Supplementary Information Section 4). 

\noindent
We now discuss how the width of the QHPs evolves with the carrier density for the High$_\text{mob}$ and Low$_\text{mob}$ regions. We focus on $\nu=6$, since this is the only plateau that was fully observed for all storage times in the range of magnetic fields accessible in our experiments. In Fig.~\ref{Fig3}b, the plateau width for each region is presented as a function of the carrier density in the system. The plateau width shows a clear linear dependence on the carrier density, growing as the electron density increases for both regions (note that the time evolution is the opposite for each region, as indicated by the arrow directions in Fig.~\ref{Fig3}). For the High$_\text{mob}$ region, the electron density increases with time and so does the plateau extension; from an initial $\Delta B=1\unit{T}$ to almost triple after 300 days plus the UV treatment. In contrast, storage time reduces the number of carriers in the Low$_\text{mob}$ region, and the QHP for $\nu=6$ shrinks from $2\unit{T}$ until disappearing at $n_c^\text{Low}\approx 2.3\cdot 10^{11} \unit{cm}^{-2}$. For comparison, the plateau in the high mobility region disappears at a larger critical carrier density of $n_c^\text{High}\approx3\cdot 10^{11} \unit{cm}^{-2}$ (obtained by extrapolating the experimental points).

\begin{figure*}[t]
    \centering
    \includegraphics[width=1\textwidth]{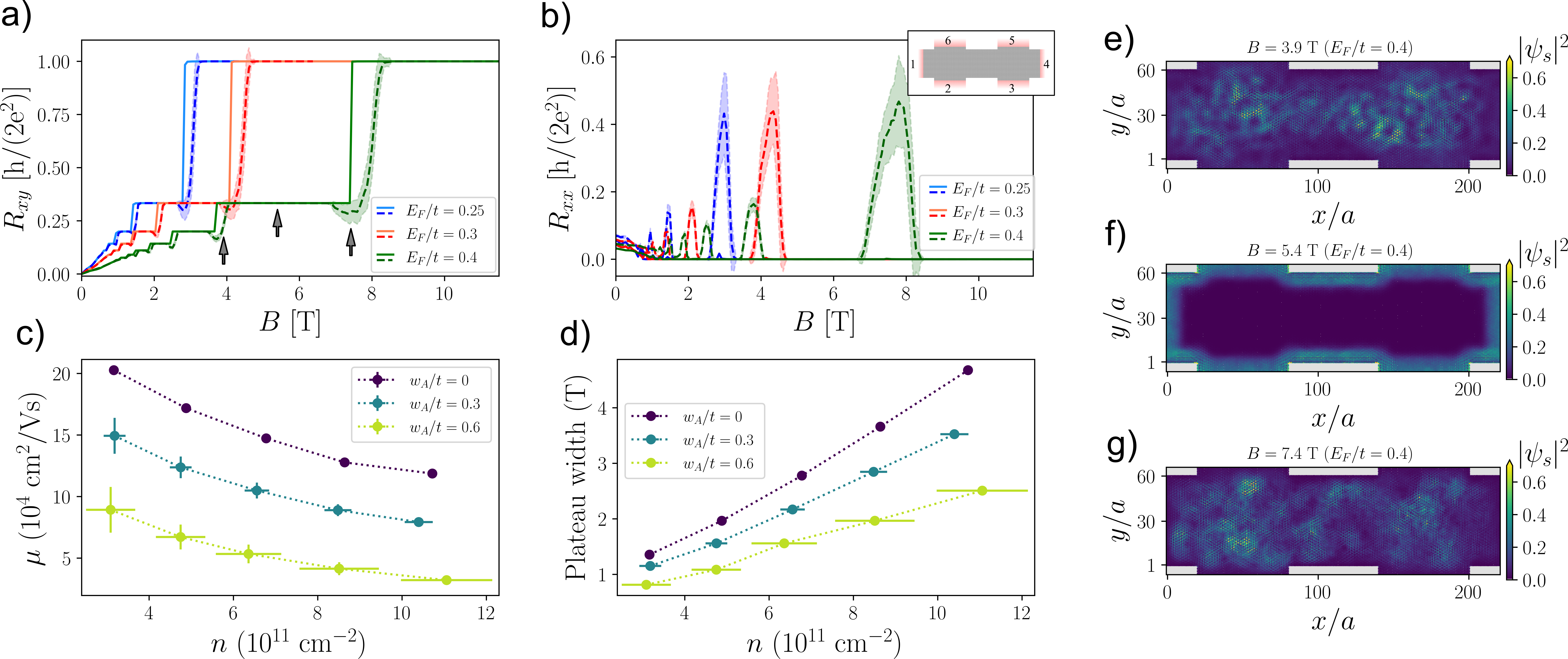}    \caption{a) Transversal resistance $R_{xy}$ as a function of the magnetic field $B$ for three different values of the chemical potential $E_F$. The solid lines correspond to the case without disorder and the dashed lines show the mean value of the resistance over 100 realizations for $\omega_A/t=0.3$, with the shadowed areas indicating the standard deviation over the disorder realizations. b) Longitudinal resistance for the same parameters as in a). c) Mobility $\mu$ and d) plateau width as a function of carrier density $n$ for $\nu=6$ and three different disorder strengths (the error bars represent the propagation of the standard deviation of the disorder realizations). e), f) and g) Scattering wavefunction density distributions $|\psi_s|^2$ for 3 different magnetic fields indicated by arrows in a).}
    \label{Fig4}
\end{figure*}

\noindent

\noindent
Even though both regions present similar carrier densities along different stages of storage, the measured QHP widths are different. The high mobility region displays, overall, broader plateaus than the low mobility one. To obtain further microscopic insight on the impact of disorder and carrier density in the QHP width, we numerically compute the transport properties of a graphene Hall bar described by a scaled tight-binding model \cite{Liu_2015} (see Supplementary Information Section 1). Disorder in the system is modeled by an Anderson on-site energy that takes random values distributed along $\epsilon_i \in [-\omega_A/2,\omega_A/2]$, where $\omega_A$ quantifies the disorder strength \cite{Roche_2012}. The Anderson model can describe qualitatively the effect of uncorrelated and short-range disorder in the samples, without resorting to more sophisticated ab-initio techniques. The results from the microscopic calculations, averaged over 100 disorder realizations (see Supplementary Information) , are collected in Fig.~\ref{Fig4}.\\
\noindent
In Fig.~\ref{Fig4}a, the transversal resistance is represented for three chemical potentials $E_F$ as a function of the applied magnetic field $B$. The ideal cases without disorder are plotted with solid lines and show perfectly quantized plateaus with sharp transitions. The cases with Anderson disorder are represented with dashed lines, the shadowed areas corresponding to the standard deviation of the disorder realizations with respect to the mean value. A disordered potential landscape destroys the quantization near the edges of each plateau as the electrons can get from one side of the sample to the other by impurity localized states. As a result, there is a smooth transition between plateaus rather than the steps shown for the ideal case. The loss of the quantization between the plateaus also impacts the longitudinal resistance, 
represented in Fig.~\ref{Fig4}b. In fact, $R_{xx}$ develops large non-zero peaks for magnetic fields corresponding to the $R_{xy}$ plateau transitions, experimentally corresponding to the SdH oscillations. The numerical calculations indicate that the plateaus shift towards higher magnetic fields when the chemical potential is increased, as already reported in the literature \cite{Krstaji__2013} and in accordance with the increments in the carrier density observed in the experiments. The simulations are thus consistent with the experimental results for the High$_\text{mob}$ vacuum stored (Low$_\text{mob}$ air stored) sample where the carrier density increased (decreased) due to oxygen desorption (absorption). 

\noindent
Figure~\ref{Fig4}c shows the relation between mobility and carrier density from the numerical simulations. The case without disorder (purple dots) displays a Drude-like behavior where the mobility is reduced with augmenting the carrier density as $\mu\propto 1/\sqrt{n}$. The effect of disorder, quantified by the Anderson disorder potential $\omega_A$, is to reduce the magnitude of the mobility. This reduction can be close to an order of magnitude at high carrier densities. The presence of disorder also induces a small change in the slope of the mobility-density ratio, which reduces slightly at higher disorder strength. The effect of disorder in our numerical simulations is thus consistent with the differences between High$_\text{mob}$ and Low$_\text{mob}$ regions shown in Fig.~\ref{Fig3}a. 
We now focus on the experimentally relevant plateau for the filling factor $\nu=6$ and resistance $R_{xy}=R_0/3$, with $R_0=h/(2e^2)$. We characterize the plateau width by the quantity $\Delta B$ defined by the difference between the minimum and maximum value of the magnetic field where the standard deviation of the transversal resistance, $\sigma(R_{xy})$, is bigger than a given threshold [$\sigma(R_{xy})> 10^{-3}R_0/3$]. For a fixed disorder strength in the system, $\Delta B$ increases with the carrier density $n$ almost linearly at high densities, see Fig.~\ref{Fig4}d. The presence of disorder reduces the plateau width for all carrier densities. At low carrier density, the calculations show a nonlinearity of the plateau width with the carrier density. Still, extrapolating from low or high carrier densities, we observe that the intercept (related to the critical carrier density observed in the experiments) increases with diminishing disorder strength. These calculations explain why the plateau width for the Low$_\text{mob}$ region stored in air shrinks with time, as oxygen absorption reduces carrier density. Conversely, in the vacuum-stored High$_\text{mob}$ region the carrier density increases, resulting in a broader plateau. Moreover, the differences in disorder between both regions provide an explanation for the observed overall differences in the plateau widths for the studied carrier density ranges. Therefore, our numerical calculations capture the experimental observation of an enlargement of the plateau width with increasing carrier density and a reduction of the quantized region due to an increase in disorder, see Figs.~\ref{Fig3}b and~\ref{Fig4}d. 

\noindent
Finally, we analyze the microscopic effect of disorder on the electronic transport by plotting the density of the space-resolved scattering wavefunction for some relevant cases. The scattering wavefunction $\psi_S$ is a valuable tool for examining the spatial distribution of the electronic density of the scattering states \cite{Groth_2014, Santos_2019}. The scattering wavefunction density, defined as $|\psi_S|^2$, is plotted in Figs.~\ref{Fig4}e, f and g for the three magnetic field values marked in Fig.~\ref{Fig4}a, and for a disorder configuration corresponding to $\omega_A/t=0.3$ and chemical potential $E_F/t=0.4$. In panels e and g, the magnetic flux values considered correspond to a transition point between plateaus. Consequently, $|\psi_S|^2$ is affected by the landscape of the Anderson disorder potential and appears delocalized in the central part of the scattering region, indicating electron scattering from edge to edge leading to non-quantized transversal resistance values. By contrast, Fig.~\ref{Fig4}f shows the density of the scattering wavefunction $|\psi_S|^2$ for a magnetic field corresponding to a perfectly quantized transversal resistance value. The density now peaks around the edges of the sample, displaying a perfectly transmitting edge state which is resilient to the disorder. 

\section*{IV. Conclusions} \label{Res}
We have studied the evolution of the QHE characteristics in epitaxial graphene for different carrier densities and different storage conditions. By means of electrical transport measurements, the mobility and the carrier density for two different regions were obtained. For the high mobility region, by storing the sample in vacuum an increase in the carrier density was observed together with a major decrease in the mobility. The QHPs shift towards higher magnetic fields due to the LL degeneracy. By UV illumination, we can conclude that the induced changes are driven by molecular oxygen being desorbed from the surface with a consequent increase in the carrier density due to its electronegativity. For the low mobility region, the evolution of the carrier density and QHE characteristics were studied as the sample was kept in air. The obtained results showed that after around 350 days, the carrier density decreased almost a factor of 10, the mobility increased up to 200$\%$ of the initial value and the QHP shifted towards lower magnetic fields. Moreover, the mobility difference within the same graphene sample can be attributed to a difference in local impurity density. We observe an increase in plateau width with increasing carrier density, attributed to larger energy spacing between LLs.  Nevertheless, a difference in plateau width between the high and low mobility regions was observed for similar carrier densities. Transport simulations based on a scaled tight-binding model with disorder show a relation between mobility and carrier density analogous to the experiments. The effect of disorder in the simulations is to decrease the magnitude of the mobility and the plateau width by means of electron scattering from the edge to the bulk states. 

\noindent
This study contributes to the understanding of the effects of disorder in the QH regime in graphene. The observation of the evolution of QHP width for different carrier densities and their dependence on the intrinsic disorder of the sample highlights the important role played by these parameters in the Quantum Hall state. This is not only relevant for an in-depth understanding but also for applications of graphene, such as metrology, where understanding the role of disorder is a major task in the development of future stable and reliable resistance standards operational in relaxed experimental conditions.

\section*{Acknowledgements} \label{Res}
We thank Ming-Hao Liu and Rafael A. Molina for valuable discussions. 
This work was supported by the Spanish Ministry for Science and Innovation (MCIN) under projects PGC2018-098613-B-C21 (SporQuMat) PID2021-122980OB-C52 (ECoSOx-ECLIPSE), PID2021-124585NB-C33 and EMPIR 20FUN03 COMET project. IMDEA Nanociencia acknowledges support from the “Severo Ochoa” Programme for Centres of Excellence in R$\&$D (Grants SEV-2016-0686 and CEX2020-001039-S). I.F.C holds an FPI fellowship from AEI-MCIN (PRE2020-092625). Y.B. and P.B. acknowledge support by the Spanish CM “Talento Program” project No. 2019-T1/IND-14088 and No. 2023-5A/IND-28927, the Agencia Estatal de Investigación project No. PID2020-117992GA-I00 and No. CNS2022-135950 and through the “María de Maeztu” Programme for Units of Excellence in R$\&$D (CEX2023-001316-M). A.J.-P., F.G. and P.P. acknowledge support from NOVMOMAT, project PID2022-142162NB-I00 funded by MICIU/AEI/10.13039/501100011033 and by FEDER, UE as well as financial support through the (MAD2D-CM)-MRR MATERIALES AVANZADOS-IMDEA-NC. E.C. acknowledges funding from Spanish Project - CNS2022-136203 - Consolidación investigadora.

\section*{ORCID Ids} \label{Res}
\noindent
Ignacio Figueruelo-Campanero: \href{https://orcid.org/0000-0001-5144-9375}{0000-0001-5144-9375}\\
Yuriko Baba: \href{https://orcid.org/0000-0003-0647-3477}{0000-0003-0647-3477}\\
Alejandro Jimeno-Pozo: \href{https://orcid.org/0000-0002-6270-7066}{0000-0002-6270-7066}\\
Julia García-Pérez: \href{https://orcid.org/0000-0002-4527-6145}{0000-0002-4527-6145}\\
Elvira M. González: \href{https://orcid.org/0000-0001-9360-3596}{0000-0001-9360-3596}\\
Rodolfo Miranda: \href{https://orcid.org/0000-0002-1064-6724}{0000-0002-1064-6724}\\
Francisco Guinea: \href{https://orcid.org/0000-0001-5915-5427}{0000-0001-5915-5427}\\
Enrique Cánovas:\href{https://orcid.org/0000-0003-1021-4929}{0000-0003-1021-4929}\\
Daniel Granados: \href{https://orcid.org/0000-0001-7708-9080}{0000-0001-7708-9080}\\
Pierre Pantaleón:\href{https://orcid.org/0000-0003-1709-7868}{0000-0003-1709-7868}\\
Pablo Burset: \href{https://orcid.org/0000-0001-5726-0485}{0000-0001-5726-0485}\\
Mariela Menghini \href{https://orcid.org/0000-0002-1744-798X}{0000-0002-1744-798X}\\
\bibliographystyle{unsrt}
\bibliography{references}

\end{multicols}
\end{document}